\documentclass[aps,pra,amsmath,amssymb]{revtex4}
\usepackage{graphicx}
\usepackage{dcolumn}
\usepackage{bm}

\begin{document}

\title{Testing of spin ordering Hamiltonian with ultracold atoms in optical lattices}
\author{G. E. Akpojotor}
\affiliation{Max-Planck-Institut f\"ur Physik komplexer Systeme,
N\"othnitzer Stra$\beta$e 38, D-01187 Dresden, Germany\\ Physics Department, Delta State University, 331001 Abraka, Nigeria}
\author{W. Li}
\affiliation{Max-Planck-Institut f\"ur Physik komplexer Systeme,
N\"othnitzer Stra$\beta$e 38, D-01187 Dresden, Germany}

\date{\today}
\keywords{spin interaction, ultracold atoms, optical lattice}
\begin{abstract}
Laser cooling and trapping are now widely used in atomic physics laboratory. Interestingly, cold atoms in optical lattices are now used in advanced research to mimic phenomena in condensed matter physics and also as a test laboratory for the models of these phenomena. It follows then that it is now possible and necessary to advance the atomic physics laboratory by including the use of ultracold atoms in optical lattices for instructional contents of phenomena in condensed matter physics. In this paper, we have proposed how to introduce into the atomic physics laboratory the study of quantum magnetism with cold atoms in a double well optical lattice. In particular, we demonstrates how to compare the theoretical parameters of a spin Hamiltonian model with those extracted from spin ordering experiment.

\end{abstract}
 \maketitle

\section{Introduction}
Cold atoms in optical lattices is the application of two formerly distinct aspects of physics: quantum gases from atomic physics \cite{Foot05} and laser theory from quantum optics \cite{Fox06}. The optical lattices are artificial crystals of light, that is, a periodic  intensity pattern formed by interference of two or more laser beams. As an insight, a pair of these laser beams in opposite directions (that is, two orthogonal standing waves with orthogonal polarization) will give a one-dimensional (1D) lattice, two pairs in two opposite directions can be used to create a 2D lattice and a similar three pairs in opposite directions will give a 3D lattice.  Atoms can be cooled and trapped in these optical lattices.  Thus in simple form, an optical lattice looks effectively like an egg carton where the atoms, like eggs, can be be arranged one per well to form a crystal of quantum matter \cite{Bloch08}.  Though the cold atoms in optical lattices was initially used to investigate quantum behaviour such as Bloch oscillations, Wannier-Stark ladders and tunneling phenomena usually associated with crystals in a crystalline solid \cite{Dahan96,Wilkinson96}, it is the theoretical proposal \cite{Jaksch89} and consequent experimental realization \cite{Greiner02} of the superfluid to Mott insulator (SF-MI) transition which is an important phenomenon in condensed matter physics that has given rise to the possibility of using it as a test laboratory for phenomena in condensed matter physics. The success of the SF-MI transition in turn emanates from the  laboratory observation of Bose Einstein condensation (BEC). The history of BEC began in 1924 when Satyendra Nath Bose first gave the rules governing the behaviour of photon which is the commonest boson. Excited by this work, Einstein in the same year extended the rules to other bosons and thereby gave birth to the Bose-Einstein distribution (BED) \cite{Libboff92}. While doing this, Einstein found that not only is it possible for two bosons to share the same quantum state at the same time, but that they actually prefer doing so. He therefore predicted that when the temperature goes down, almost all the particles in a bosonic system would congregate in the ground state even at a finite temperature. It is this physical state that is called Bose-Einstein condensation. Thus it has always been considered a consequence of quantum effects from statistical mechanics in many textbooks as the phase transition  is achieved without interactions \cite{Libboff92,Stringari05}. The Einstein's prediction, however, was considered a mathematical artifact for sometime until Fritz London in 1938 while investigating superfluid liquid helium, realized that the phase transition could be accounted for in terms of BEC. This analysis, however, suffered a major set back because the helium atoms in the liquid interacted quite strongly. This was why scientists had to move ahead in search of BEC in less complicated systems that would be close to the free boson gas model. Fortunately, the breakthrough came in 1995 when the first BEC was observed in rubidium atoms and this was followed by similar observations in some other cold alkali atoms such as those of lithium and sodium (see more details in Ref. (\cite{Weiman96} and a guideline to the literature of BEC in dilute gases in Ref. \cite{Hall03}.\\
As stated above, the observation of BEC led to the observation of (SF-MI) transition and thereby open the possibility to investigate various phenomena in condensed matter physics by mimicking them with ultracold atoms in optical lattices. This possibility has led to a deluge of studies (see  \cite{Lewenstein07} for a recent review) as it brings together atomic physicists, quantum opticians and condensed matter physicists.  One draw back is that even when there have been theoretical papers investigating these phenomena with fermionic cold atoms \cite{Rey07}, cold bosons are used in the actual experiments \cite{Trotzky08} for testing spin ordering. This has been overcome by the recent observation of the MI with fermonic atoms \cite{Jordan08}.  It follows then that the possibility to use cold atoms in optical lattices as a test laboratory for condensed matter physics is no longer a speculative physics. Rather, it has become an aspect of physics with its own methods and approaches. Therefore, it  has reach a stage when it should start having some introductory impact on our curriculum, possibly as applications of optics, atomic physics and simulation of spin ordering Hamiltonians \cite{Tsai08} in condensed matter physics.  The purpose of this current study is to present a pedagogical study of investigating spin ordering in an isolated double - well - type potential (simply  double well (DW)) which can be adopted for instructional purposes.  For the DW is one of the simplest  experimental set ups of optical lattices  to study spin Hamiltonians \cite{Folling07}.  This is because the system can be completely controlled and measured in an arbitrary two-spin basis by dynamically changing the lattice parameters \cite{Rey07}. On the theoretical side, the DW can be considered as two localized spatial modes separated by a barrier and consequently be investigated as a two-mode approximation \cite{Jaksch89}.

\section{the double-well superlattice}

The DW is a 1D  optical lattice in which the transverse directions are in strong confinement and thus the motions of an atom in these directions are frozen out. To create the DW, we start with a standing wave of period $d$ (long lattice) so that the potential seen by the atoms trapped in it is
\begin{equation}
\label{opotentil}
V_1(x)=V_1\text{cos}^2(\pi x/d)
\end{equation}
where $V_1$ is the lattice depth, which is a key parameter for a special lattice potential.\\
Next we superpose a second standing wave with period $d/2$ and depth $V_2$ (short lattice) on the first one as in Fig. 1(a). This will lead to a symmetric double-well superlattice (Fig 1c) with a total optical lattice \cite{Anderlini07,Trotzky08}
\begin{equation}
\label{topotential}
V(x)=V_1(x)+V_2\text{cos}^2[2\pi x/d].
\end{equation}
 The configuration and varying of the parameter space (i.e. various parameters) of a Hamiltonian to be tested in this superlattice is achieved by  manipulating and controlling the depths of the short and long lattices.  For example, by increasing the lattice depth of long-lattice $V_1$, we could reach from superfluid to Mott-insulator regime, which is convenient for studying the few particles phenomena in a local double-well cell. And the barrier height of the double-well is controlled by the lattice depth of short-lattice, $V_2$. The effective double-well is reached if $V_1>4V_2$. Otherwise the minimal points of the optical lattice are the bottom of the long-lattice. This could be seen clearly from Eq. (\ref{topotential}) after we expand the $\text{cos}(2\pi x/d)$ term,
\begin{equation}
\label{minimal}
V(x)=4V_2\left[\text{cos}^2\left(\frac{\pi x}{d}\right)+\frac{1}{2}\left(\frac{V_1}{4V_2}-1\right)\right]^2+\frac{V_1}{4}\left(2-\frac{V_1}{4V_2}\right),
\end{equation}
so that $\text{cos}^2 z\geq 0$ results in $V(x)_{min}=V_2$, that is, the minimum is one of the  long-lattice with a  lift of $V_2$. On the other hand, if $V_1<4V_2$, an effective double-well is created. The minimum can be found at $\text{cos}^2(\pi x/d)=[1-V_1/(4V_2)]/2$. Making a power series expansion around the potential minimum, then a single atom of mass $m$ trapped initially in any of the well will freely tunneling back and forth with the oscillation frequency of
\begin{equation}
\label{frequency}
\omega=\frac{\pi}{d}\sqrt{\frac{(16V_2^2-V_1^2)}{2mV_2}}.
\end{equation}
Thus the frequency depends on not only the lattice depths $V_1$ and $V_2$, but also on the lattice spacing $d$. Usually, the small lattice spacing $d$ is preferred as it leads to a large frequency though this could also be restricted by changing the ratio $V_1/4V_2$. This preference also lead to the use of the recoil energy of the short lattice as the unit of the depths of the optical lattice
\begin{equation}
\label{recoil}
E_r=\frac{h^2}{2m\lambda^2}
\end{equation}
where $\lambda$ is the wave length of the short lattice.\\
For example, in the experiment \cite{Trotzky08}, the depth of the long lattice is $10E_r$ while the depth of the short lattice is about  $6E_r<V_2<44E_r$. This gives the oscillation frequency in a range
\begin{equation}
\label{frange}
\frac{\pi}{d}\sqrt{\frac{119E_r}{3m}}<\omega<\frac{\pi}{d}\sqrt{\frac{7719E_r}{22m}}.
\end{equation}
For $^{87}$Rb and $d=\lambda=765\text{nm}$, we can get $E_r=2.596\times10^{-30}J$, which gives $109kHz<\lambda<326kHz$. Lets define the harmonic oscillator length $a_o=\sqrt{\hbar/m\omega}$, then we can readily get $118.6\text{nm}<a_o<204.5\text{nm}$. Comparing to the period of short lattice $d/2=382.5\text{nm}$, it implies the ground state wave function is rather localized, which ensures the validity of the two-mode approximation.\\
Finally, it is pertinent to describe how to create an asymmetric DW (Fig 1d). The potential bias or the tilt $\Delta$ of the double-well is introduced by changing the relative phase of the two potentials (i.e. short and long lattices) and this can be realized by applying a magnetic field gradient of $ B'$ \cite{Trotzky08}. Consequently, tuning $B'$ gives the potential difference between the two potential minima of the DW. We can realize the adiabatic and diabatic operations on the tilt of the DW by controlling the increasing speed of $B'$ \cite{Sebby06}.
\begin{figure}
\centering
\includegraphics*[width=0.40\columnwidth]{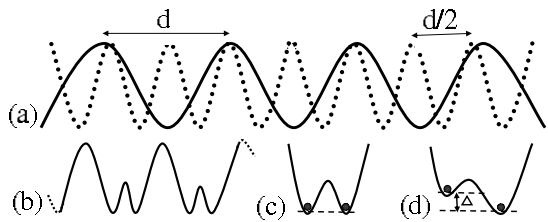}
\caption{(a) Two standing waves in opposite directions and with periods $d$ and $d/2$ resulting in (b) a chain of double wells from which we can study (c) an isolated symmetric double well or (d) asymmetric double well}.
\label{double well}
\end{figure}

\section{two-mode model with only on-site interaction}

Within the above consideration and assuming atoms confined in an isolated DW, we reach the two-mode approximation represented as a two-site version of the Hubbard model \cite{Essler05}
\begin{equation}
H_0=\sum_{\sigma=\uparrow,\downarrow}\left[-J(a_{\sigma L}^{\dagger}a_{\bar{\sigma} R}+a_{\sigma R}^{\dagger}a_{\bar{\sigma} L})-\frac{1}{2}\Delta(n_{\sigma L}-n_{\sigma R})\right]+U(n_{\uparrow L}n_{\downarrow L}+n_{\uparrow R}n_{\downarrow R})
\label{h1}
\end{equation}
where $a_{\sigma L,R}^{\dagger}(a_{\bar{\sigma} L,R})$ is the creation operator (annihilation operator) for an atom with spin $\sigma (\bar{\sigma})= \uparrow(\downarrow),\downarrow(\uparrow)$, $n_{\sigma,L,R}$ is the corresponding number operator, $J$ (both J and t are used in the literature though the cold matter community seems to prefer J) describes the tunneling rate between the two wells, $\Delta$ is the potential bias for the double-well and $U$ is the two-body interaction when two atoms occupy the same site.\\
Basically, the Hubbard model is a single band lattice model supporting a single atomic state which can hold up to two particles. If we consider that these particles have two internal spins, $\sigma (\bar{\sigma})$, then the Hamiltonian will consist of the superposition of six Fock basis states \cite{Parker05,Avelar05} denoted by $|\uparrow \downarrow,0\rangle, |\uparrow,\downarrow\rangle, |\downarrow,\uparrow\rangle, |0,\uparrow \downarrow\rangle, |\uparrow,\uparrow\rangle$ and $|\downarrow,\downarrow\rangle$, with a basis state $|l,r \rangle$  denoting the $l = left$ and $r = right$ wells. The use of  the superposition principle which is a fundamental concept in quantum theory \cite{Bransdon92} is consistent with both the model Hamiltonian and experiment in which the time evolution of the initial states produces coherent superposition of states. Thus the wavefunction of the system will be the superposition of all possible states
\begin{equation}
\label{wavefunction1}
\vert\psi\rangle = \sum_{i = l,r} \vert\phi_{ii}\rangle_{\sigma,\bar{\sigma}}^{\pm} + \sum_{<i,j> =l,r} \vert\phi_{ij}\rangle_{\sigma,\bar{\sigma}}^{\pm} + \sum_{<i,j> =l,r} \vert\phi_{ij}\rangle_{\sigma,\sigma}^{\pm}.
\end{equation}
where for convenience, $\vert\phi_{ii}\rangle_{\sigma,\bar{\sigma}}^{\pm} =\vert t\rangle/\vert s\rangle_{\sigma,\bar{\sigma}}^{\pm} = {|\uparrow \downarrow,0\rangle \pm |0,\downarrow\uparrow\rangle}/\sqrt{2}$, $\vert\phi_{ij}\rangle_{\sigma,\bar{\sigma}}^{\pm} = \vert t\rangle/\vert s\rangle_{\sigma,\bar{\sigma}}^{\pm} ={|\uparrow, \downarrow\rangle \pm |\downarrow,\uparrow\rangle}/\sqrt{2}$ and $\vert\phi_{ij}\rangle_{\sigma,\sigma}^{\pm} = \vert t\rangle_{\sigma,\sigma}^{\pm} ={|\uparrow, \uparrow\rangle \pm |\downarrow,\downarrow\rangle}/\sqrt{2}$, with i, j denoting the sites while $\vert\ s\rangle$ and $\vert\ t\rangle$ denote singlet and triplet states respectively. From basic physics, the orientation of the two spins in a state can either be singlet $\vert\ s\rangle$  if $S = 0$ or triplet $\vert\ t\rangle$ if $S = 1, 0, -1$ \cite{Bransdon92, Petukhov92, Sachdev90}.

\subsection{eigenenergies and eigenstates at $\Delta=0$}
By tuning the potential bias $\Delta=0$, we can obtain all the eigenenergies and corresponding eigenstates analytically. This is achieved by directly diagonalizing the Hamiltonian in Eq. (\ref{h1}) to obtain eigenenergies and eigenstates as shown in Table (\ref{tabeigen1}). In the weak interacting case, $U << J$, the state $|\uparrow\downarrow,0\rangle$ and state $|0,\uparrow\downarrow\rangle$ have lower energy so that the doubly occupied singlet state, $\vert\phi_{ii}\rangle_{\sigma,\bar{\sigma}}^{-} = \vert s\rangle$ will be the ground state of the system. This can be considered as the signature for a superfluid state for bosonic atom in a double well. However, the strong interaction regime is more interesting to study for spin ordering. In this regime, $U >> J$, the ground state will be singly  occupied as the large atomic repulsion energetically suppress the double occupancy. Here it is the $\vert\phi_{ij}\rangle_{\sigma,\bar{\sigma}}^{\pm}$ that are occupied while the $\vert\phi_{ij}\rangle_{\sigma,\sigma}^{\pm}$ are unpopulated \cite{Rey07,Sachdev90}. The populated $\vert\ s\rangle$  and $\vert\ t\rangle$ of $\vert\phi_{ij}\rangle_{\sigma,\bar{\sigma}}^{\pm}$ are nearly degenerate because the energy difference between them is  about $4J^2/U$, which is a small quantity. However, when $J\approx0$, the ground state approaches $\vert\ s\rangle$ while the first excited state is $\vert\ t\rangle$. If we prepare the initial state as antiferromagnetic, $|\uparrow,\downarrow\rangle=(|t\rangle+|s\rangle)/\sqrt{2}$, the dynamical evolution involves two frequencies \cite{Trotzky08}
\begin{equation}
\hbar\omega_{1,2}=\frac{\sqrt{16J^2+U^2}\pm U}{2}.
\label{frequency}
\end{equation}
From the above frequencies, one could get the tunneling rate $J = 1/2\hbar\sqrt{\omega_{1}\omega_{2}}$ and the interaction strength $U=\hbar(\omega_1-\omega_2)$ respectively. These two frequencies can be obtained from exerimental data and then used to test the validity of the simple two-mode model. This has been done for bosonic atoms in experiment \cite{Trotzky08}. The extension to fermionic atoms may be different but the eigenenergies and corresponding eigenstates are the same, which means we can get similar dynamics as long as the interaction between atoms satisfy $U>0$ \cite{Folling07}. On the other hand, the interaction of fermion could be also attractive generally. It is interesting to identify the ground state in this situation. Table (\ref{tabeigen1}) works here too.
\begin{table}[h]
\caption{Eigenstates and Eigenenergies of Hamiltonian (\ref{h1})}
\centering
\begin{tabular}{ccccccc}
\hline
\cline{2-7}
Eigenenergy &$|\downarrow,\downarrow\rangle$&  $|\uparrow \downarrow,0\rangle$ & $|\uparrow, \downarrow\rangle$ & $|\downarrow,\uparrow\rangle$ & $|0,\uparrow \downarrow\rangle$  & $|\uparrow,\uparrow\rangle$\\
\hline
$\frac{U-\sqrt{16J^2+U^2}}{2}$ &0& 1  & $\frac{U+\sqrt{16J^2+U^2}}{4J}$  & $-\frac{U+\sqrt{16J^2+U^2}}{4J}$ &- 1&0  \\
0&0&0  & 1 & 1 &0& 0  \\
0& 1&0&0&0&0&0 \\
0& 0&0&0&0&0&1 \\
U&0&1      & 0     &  0& 1&0\\
$\frac{U+\sqrt{16J^2+U^2}}{2}$ &0&1  & $\frac{U-\sqrt{16J^2+U^2}}{4J}$  & $-\frac{U-\sqrt{16J^2+U^2}}{4J}$ & -1&0 \\
\hline
\end{tabular}
\label{tabeigen1}
\end{table}
When $U<0$, the ground state does not change too much at weak tunneling. However, the first excited state is not the triplet state $\vert\phi_{ij}\rangle_{\sigma,\bar{\sigma}}^{+} = \vert t\rangle$ anymore but the state $\vert\phi_{ii}\rangle_{\sigma,\bar{\sigma}}^{+} = \vert t\rangle$. It is interesting that the energy difference is the same $\Delta E\approx 4J^2/|U|$ although the interaction is attractive and the first excited state is changed. This analysis shows that $\vert\phi_{ii}\rangle_{\sigma,\bar{\sigma}}^{+} = \vert t\rangle$ could be involved in the dynamics if we start from the antiferromagnetic initial state $|\uparrow,\downarrow\rangle$. The two frequencies that can be observed in experiment are
\begin{equation}
\label{afreq}
\hbar \omega_{1}^{\prime}=U,\hbar\omega_{2}^{\prime}=\frac{\sqrt{16J^2+U^2}-U}{2}
\end{equation}
Here the two-body interaction strength directly relates to $\omega_{1}^{\prime}$, which can be extracted from the measured experimental data. The results show that ultracold atoms trapped in the superlattice not only could be used to simulate the phenomena in condensed matter physics, but also offer the possibility to compare the results with theoretical calculation of model Hamiltonians.

\subsection{eigenstates and eigenenergies at $\Delta\ne 0$}

\begin{figure}
\centering
\includegraphics*[width=0.32\columnwidth]{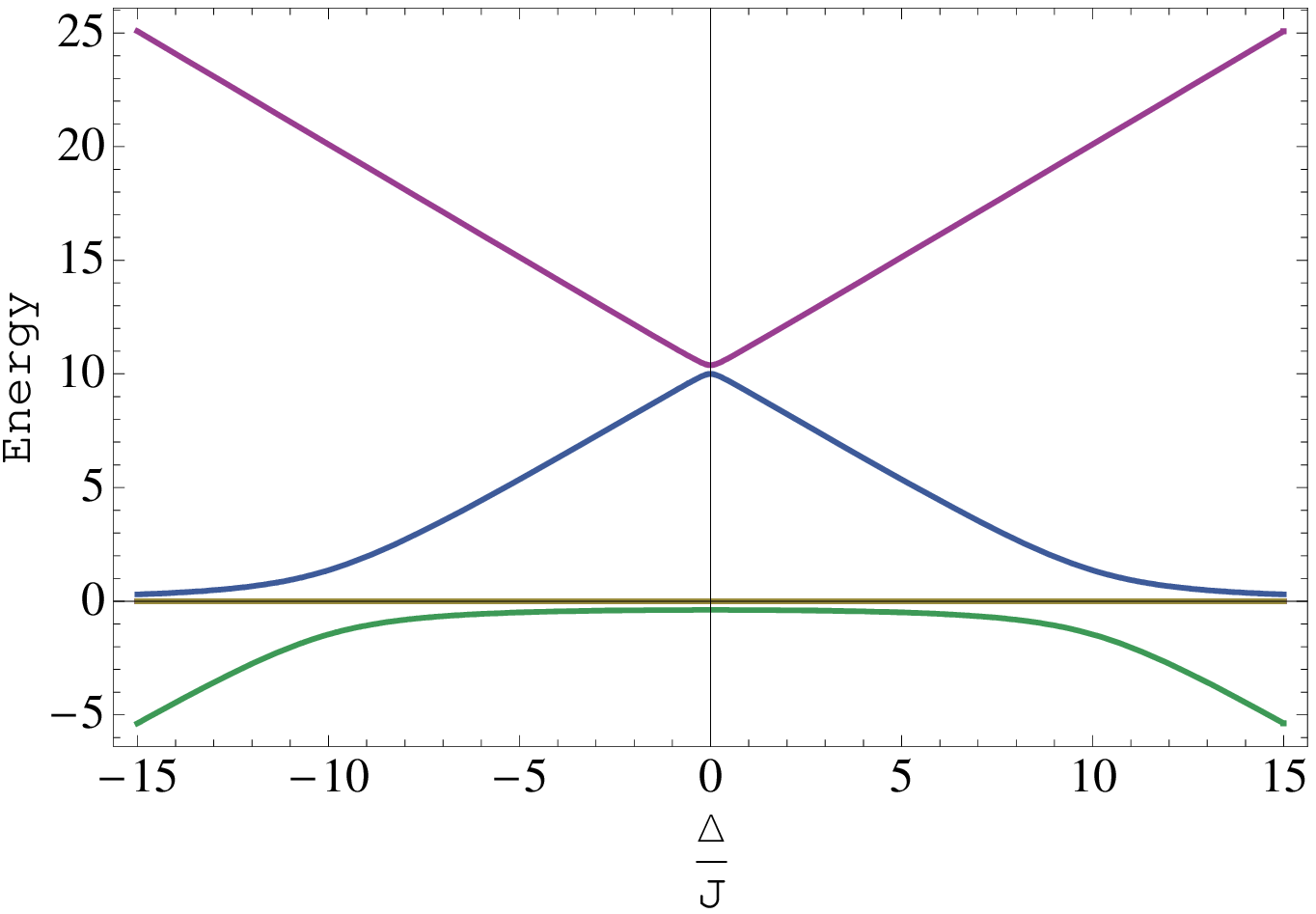}
\includegraphics*[width=0.32\columnwidth]{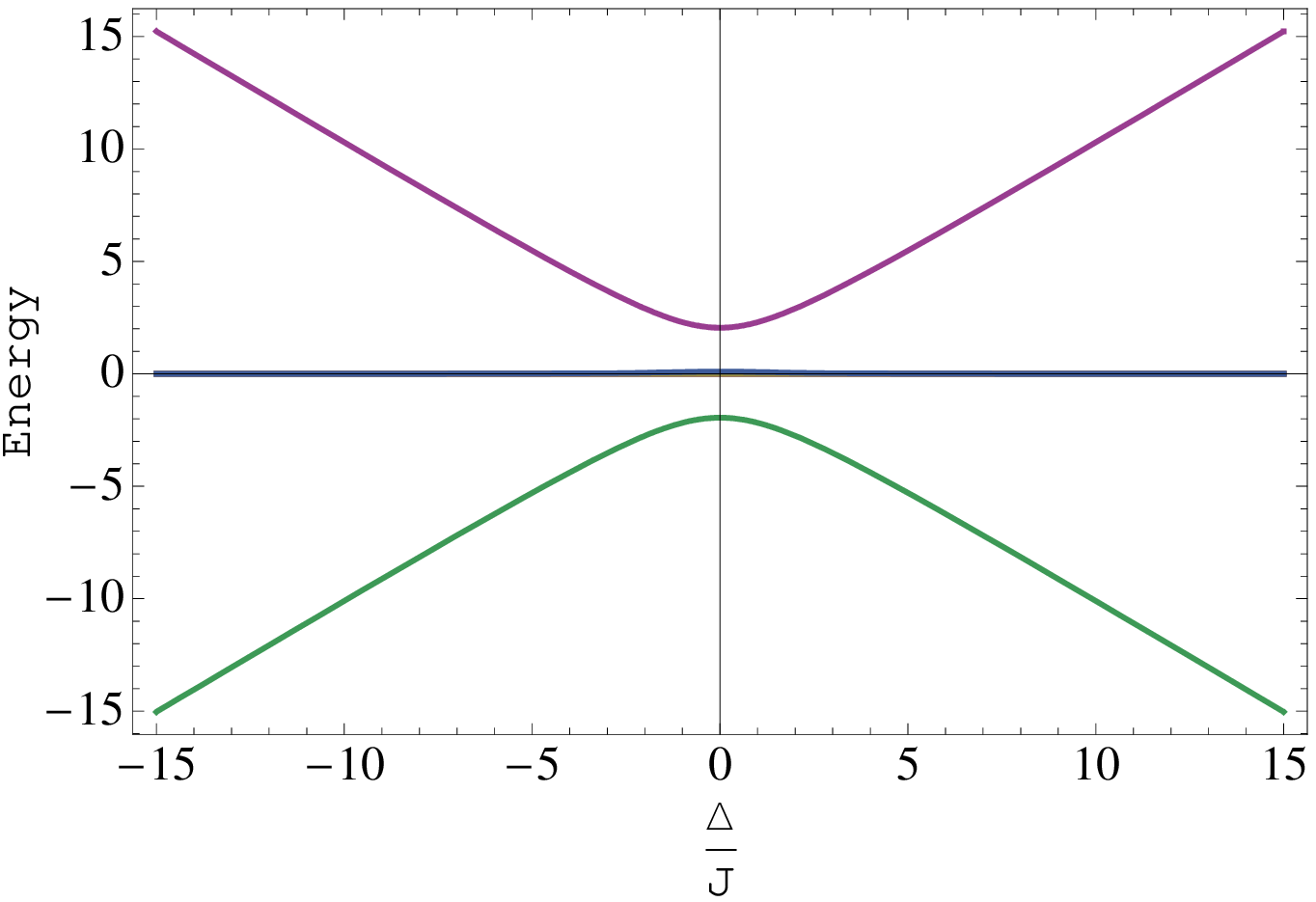}
\includegraphics*[width=0.32\columnwidth]{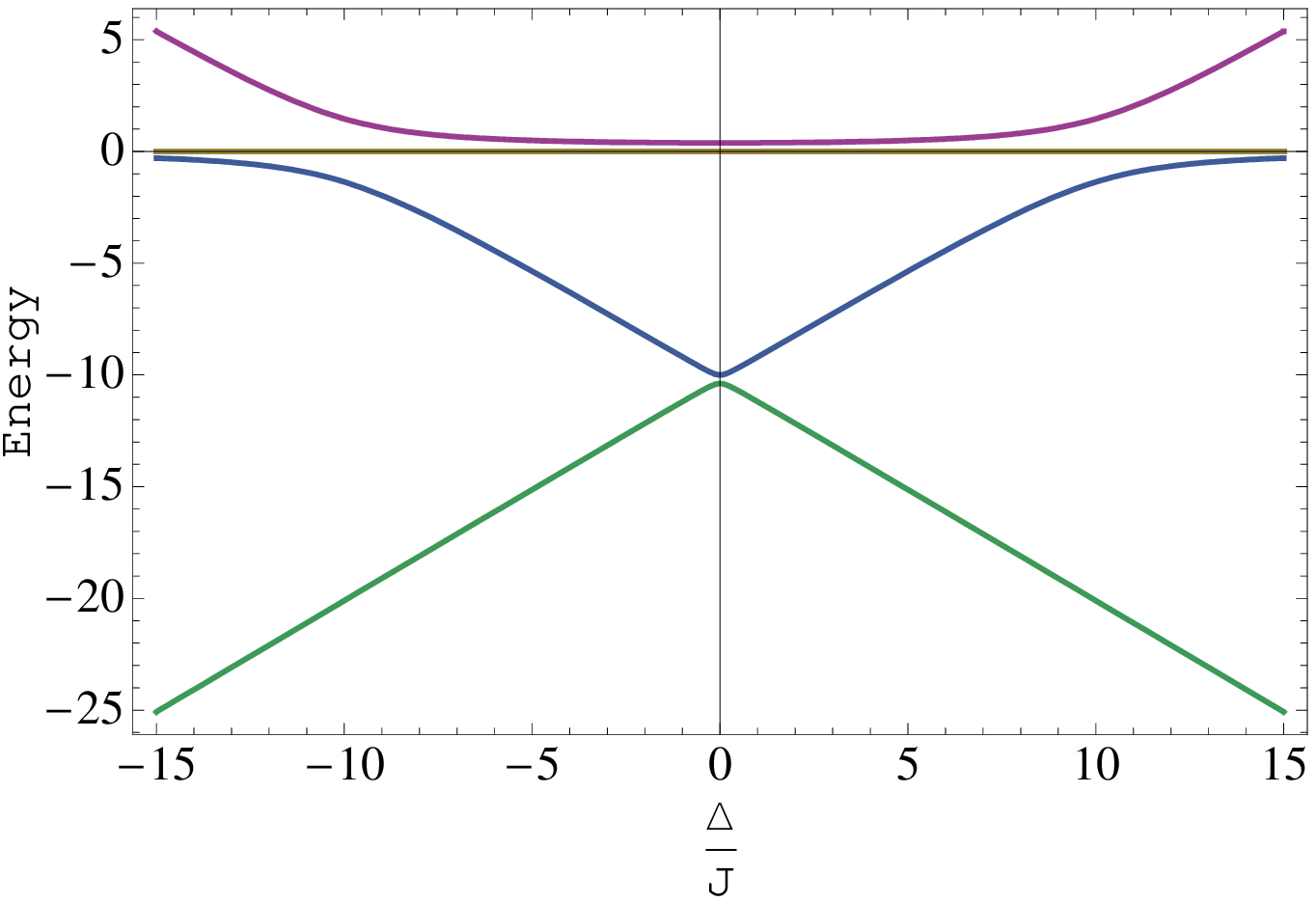}
\caption{(Color online.) Spectra of the eigenstates at different potential bias $\Delta/J$. Left, $U/J=10.0$ in the strong interaction regime. Middle $U/J=0.1$ in the weak interaction regime. Right, attractive interaction $U/J=-10$. }
\label{spectra}
\end{figure}

One of the advantages of the trapped ultracold atoms is that they can be controlled precisely and easily. By tuning the two optical lattices, we can change the bias of the double-well superlattice. The ground state of the two fermions are trapped in the same site in the large potential bias. By slowly reducing the bias, the ground state is followed adiabatically to the singlet state. In this way, we can prepare the initial states either in the state $|\uparrow\downarrow,0\rangle$ or $|0,\uparrow\downarrow\rangle$. So it is interesting to investigate how the bias influences the states.

When the potential bias $\Delta$ is included, the eigenstates and the eigenenergies have complicated expressions. Three of the eigenenergies are $0$ always. The others are the roots of the algebra equation
\begin{equation}
\label{algebra}
x^3-2Ux^2+(U^2-4J^2-\Delta^2)x+4UJ^2=0.
\end{equation}

We numerically solve the equation and plot the eigen spectra in Fig. (\ref{spectra}).  When $U\gg J$, the ground state energy is negative and modified by the presence of potential bias.  When $\Delta$ is not too big, the ground state energy is close to the one without potential bias. Also, the energy difference between the ground state and first excited state is small. At large potential bias $\Delta\gg U$, however,  the approximate ground state energy reads
\begin{equation}
\label{egdelta}
E_g=U-\sqrt{4+\Delta^2}.
\end{equation}
The two atoms are in the right well and the ground state reaches $|0,\uparrow\downarrow\rangle$. On the contrary, the system is degenerate further in the weak interaction regime $U\ll J$. If the interaction is attractive, the energy spectra is reversed. The energy difference between the  ground state and the first excited state is bigger except at $\Delta=0$, i.e. an anti-crossing appears. This is shown obviously in Fig. (\ref{spectra}). The observation from this analysis is that the  potential bias can be used to control the energy difference between the singlet and triplet states \cite{Rey07}. This is why an attempt was made in \cite{Trotzky08} to use it to drive Eq. \ref{h1} into superexchange interaction observed experimentally. The outcome, however, is that inter-well interactions  have to be included to Eq. \ref{h1} to get close to the experimental data. Thus in the next section, we will consider such an extension.



\section{long range interactions and interaction induced spin-flip}

It is obvious from the preceding section that a Hamiltonian to study spin ordering in the cold atoms in optical lattices needs to contain long range  interactions. It is important to point out that the overlapping of different electronic orbitals gives rise to the interaction between spins in condensed matter but this overlapping is very small in optical lattices \cite{Lewenstein08}. However, the possibility of the atoms to tunnel through the barrier in quantum mechanics enables the inter-site interactions \cite{Lewenstein07}.  Two natural candidates are the inter-site Coulombic interaction $V$ and exchange interaction $J_{ex}$.  Interestingly, the inclusion of these interactions as means of going beyond the standard Hubbard Hamiltonian to account for ferromagnetism in metals have been proposed \cite{Amadon96,Hirsch97}. Furthermore, we do not need a potential bias since the spin ordering is induced by these interactions. Within these considerations, the extended form of Eq. \ref{h1} with $\Delta = 0$ is
\begin{equation}
H=\sum_{\sigma=\uparrow,\downarrow}[-J(a_{\sigma L}^{\dagger}a_{\bar{\sigma} R}+a_{\sigma R}^{\dagger}a_{\bar{\sigma} L}) + U(n_{\uparrow L}n_{\downarrow L}+n_{\uparrow R}n_{\downarrow R})  +  V(n_{\uparrow L}n_{\downarrow R}+n_{\downarrow L}n_{\uparrow R})+ J_{ex}\biggl[\sum_{\sigma,\bar{\sigma}=\uparrow,\downarrow} a_{\sigma L}^{\dagger}a_{\bar{\sigma}R}^{\dagger}a_{\sigma R}a_{\bar{\sigma} L}\biggr].
\label{h2}
\end{equation}
It is then easy to obtain the ground state energy and wavefunction using the highly simplified correlated variational approach (HSCVA) in \cite{Akpojotor08}. The beauty of this pedagogical approach is that the ground state energy clearly depicts the physics of the model  as one vary the parameter space as in experiments with optical lattices. Interestingly, the method allows the decoupling of the kinetic part from the interaction parts so that we can observe the effects of including each of them to the kinetic part. Thus the combination of these two factors makes the HSCVA very suitable to investigate the spin Hamiltonian to be tested using cold atoms in optical lattices.\\
 We start with the variational ground state energy
 \begin{equation}
\label{variational}
 E_g = \frac{\langle\psi\vert H \vert\psi\rangle}{\langle\psi\vert \psi\rangle},
\end{equation}
where the H is the model Hamiltonian and the ket in the Hilbert space is the trial wave function (cf. Eq. (\ref{wavefunction1})) defined as\\
\begin{equation}
\label{wavefunction2}
\vert\psi\rangle = \sum_{i = l,r} X_{ii} \vert\phi_{ii}\rangle_{\sigma,\bar{\sigma}}^{\pm} + \sum_{<i,j> =l,r} X_{ij} \vert\phi_{ij}\rangle_{\sigma,\bar{\sigma}}^{\pm} + \sum_{<i,j> =l,r} Y_{ij} \vert\phi_{ij}\rangle_{\sigma,\bar{\sigma}}^{\pm}.
\end{equation}
The X and Y in Eq. (\ref{wavefunction2}) are the variational parameters. It is straightforward to show \cite{Akpojotor08} that Eq. (\ref{variational}) leads to a 3 x 3 blocked matrix of 2 x 2 and 1 x 1 resulting in the lowest state energies \cite{Amadon96}, $E_g/t = E_s$ for the singlet states $\vert\phi_{ii}\rangle_{\sigma,\bar{\sigma}}^{-} = \vert s\rangle$ or $\vert\phi_{ij}\rangle_{\sigma,\bar{\sigma}}^{-} = \vert s\rangle$ depending on U,
\begin{equation}
\label{singletE}
E_s=-2\biggl[\sqrt{{\biggl(\frac{U}{4J}-\frac{V}{4J}-\frac{J_{ex}}{4J}\biggr)^2}+1} - \biggl(\frac{U}{4J}+\frac{V}{4t}+\frac{J_{ex}}{4J}\biggr)\biggr]
\end{equation}
and $E_g/t = E_t$ for the triplet state $\vert\phi_{ij}\rangle_{\sigma,\sigma}^{\pm} = \vert t\rangle$,
\begin{equation}
\label{tripletE}
E_t=V-J_{ex}.
\end{equation}
The smallest of these two energies will be the ground state energy of the system. The corresponding eigenvectors are then substituted as the variational parameters in Eq. (\ref{wavefunction2}) to give the corresponding ground state wavefunctions. Thus when $E_s < E_t$, the system will be antiferromagnetic while it will be ferromagnetic otherwise. Taking into account this condition and Eqs. (\ref{singletE}) and (\ref{tripletE}), the critical value of $J_{ex}$ at which there is transition from one state to another is \\
\begin{equation}
\label{criticalJ}
\frac{J_{ex}}{4J}> \frac{1}{2}\biggl[\sqrt{{\biggl(\frac{U}{4J}-\frac{V}{4J}\biggr)^2}+1} - \biggl(\frac{U}{4J}+\frac{V}{4J}\biggr)\biggr].
\end{equation}
Now to test this Hamiltonian in a double well, we need to know how the atomic positions and spin orientations varies with the parameter space. For example, as demonstrated  in subsection (A), the ground state of the system will be a Mott insulator when the U is very strong. This generally accepted property of the half-filled standard Hubbard Hamiltonian (i.e. $V = J_{ex} = 0$) is already achieved with ultracold fermionic atoms \cite{Jordan08}.  One of the signatures of the MI state is the decrease in doubly occupied states in the ground state as $U$ is increased. This is demonstrated in Fig. 3 showing the level of occupation of the states denoted by the variational parameters of the ground state wavefunction with increase in $U$. The inclusion of $V$, however, enhances the double occupancy and is therefore expected to suppress the observation of the MI especially for low values of U. It follows then that when we switch on the $J_{ex}$, the $U$ is likely to drive the system into more singly occupied states and thereby enhancing the transition to a ferromagnetic state while the $V$ will suppress it.  This is demonstrated  in Eqs. (\ref{singletE})  - (\ref{criticalJ}) and then depicted in Fig. 4 showing the variation of the antiferromagnetic-ferromagnetic transition critical point of $J_{ex}$ with $U$ at various values of $V$.\\
The above theoretical $V$ and $J_{ex}$ can also be compared with the ones obtained from extracted data from the experiments as was done for J and U in the standard Hubbard Hamiltonian. This is by expressing the possible dynamic evolution frequencies from for the singlet states and triplet states as
\begin{equation}
\label{FreqsingletE}
\hbar\omega_{3,4}=2J\biggl[\sqrt{{\biggl(\frac{U}{4J}-\frac{V}{4J}-\frac{J_{ex}}{4J}\biggr)^2}+1} \pm \biggl(\frac{U}{4J}+\frac{V}{4J}+\frac{J_{ex}}{4J}\biggr)\biggr]
\end{equation}
\begin{equation}
\label{FreqtripletE}
\hbar\omega_5=V-J_{ex}.
\end{equation}
We see immediately that we can recover Eq. (\ref{frequency}) from Eq. (\ref{FreqsingletE}) when $V = J_{ex} = 0$. Taking into account Eqs. (\ref{frequency}), (\ref{FreqsingletE} and  (\ref{FreqtripletE}),  we can then estimate $V=\frac{\hbar}{2}[(\omega_3-\omega_4)-(\omega_1-\omega_2)+\omega_5]$ and $J_{ex}=\frac{\hbar}{2}[(\omega_3-\omega_4)-(\omega_1-\omega_2)-\omega_5]$. Thus we can also obtain the inter-site interaction parameters from the data extracted from the experiments.
\section{Conclusion}
The increasing advancement on how to prepare, manipulate and detect phenomena in condensed matter physics using cold atoms in optical lattices has reached a stage when it can be used as instructional means. The fact that laser cooling and trapping are now widely used in atomic physics laboratory \cite{Vredenbregt03} means the realization of the double wells experiment can also be achieved. The first investigation is to mimic the Mott insulator state. By extracting  $\omega_1$ and $\omega_2$ from the experiment, the experimental values of J and U can be compared with the ones from their theoretical values. The experiment can then be advanced to determine $V$ and $J_{ex}$ and then compare them with their theoretical values as depicted in Fig. 4.  Interestingly, the model Hamiltonian studied here has been proposed to account for spin ordering in transition metals. It is hoped therefore, that the testing of this extended Hubbard model can easily be compared to available data for the transition metals \cite{Hirsch97} after some refining of the approach here. This will also include extending the study to dynamic properties of the model.
\begin{figure}
\centering
\includegraphics*[width=0.28\columnwidth]{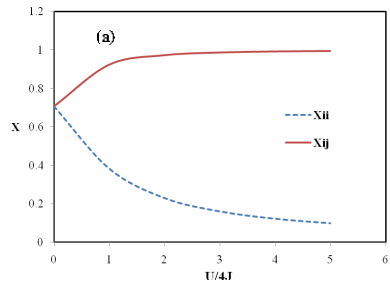}
\caption{(Color online) The decrease of the doubly occupied states $X_{ii}$  while there is increase in inter-site states $X_{ij}$ as the on-site Coulomb interactions U increases.}
\label{XvsU}
\end{figure}\begin{figure}
\centering
\includegraphics*[width=0.40\columnwidth]{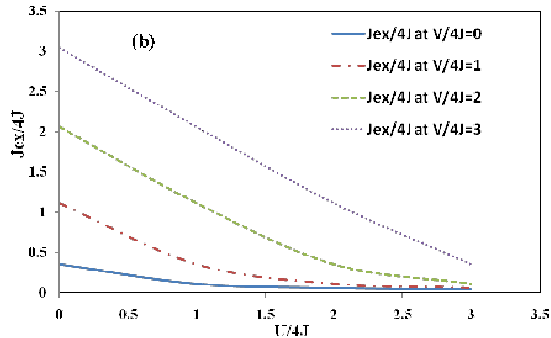}
\caption{(Color online) The variation of the antiferromagnetic-ferromagnetic transition critical point of the exchange interaction $J_{ex}$ with the on-site Coulomb interactions $U$ at various values of the inter-site Coulomb interactions $V$. A  similar graph was obtained by Ref. (\cite{Amadon96}) using different analytical method}
\label{JvsUatV}
\end{figure}
\section{Acknowledgments}
We acknowledge useful discussions with Masud Haque, Ian Spielman and Shan-Ho Tsai. GEA acknowledges partial support from AFAHOSITECH.

\end{document}